# Going beyond copper: wafer-scale synthesis of graphene on sapphire

N. Mishra[1,2], S. Forti[1], F. Fabbri[1,2], L. Martini[1], C. McAleese[3], B. Conran[3], P.R. Whelan[4,5], A. Shivayogimath[4,5], L. Buß[6], J. Falta[6], I. Aliaj[7], S. Roddaro[7,8], J. I. Flege[6,9], P. Bøggild[4,5], K.B.K. Teo[3] and C. Coletti[1,2,*]

[1]*Center for Nanotechnology Innovation @ NEST, Istituto Italiano di Tecnologia , Piazza San Silvestro 12,56127 Pisa, Italy*
[2]*Graphene Labs, Istituto Italiano di Tecnologia, Via Morego 30, 16163 Genova, Italy*
[3]*AIXTRON Ltd., Buckingway Business Park, Anderson Rd, Swavesey, Cambridge CB24 4FQ, UK*
[4]*DTU Physics, Ørsteds Plads 345C, 2800 Kgs. Lyngby, Denmark*
[5]*Center for Nanostructured Graphene (CNG), Ørsteds Plads 345C, 2800 Kgs. Lyngby, Denmark*
[6]*Institute of Solid State Physics, University of Bremen, Bremen-28334, Germany*
[7]*NEST, Scuola Normale Superiore and Istituto Nanoscienze-CNR, Piazza S. Silvestro 12, I-56127 Pisa, Italy*
[8]*Dipartimento di Fisica, Università di Pisa, Largo B. Pontecorvo 3, 56127, Pisa (PI), Italy*
[9]*Brandenburg University of Technology Cottbus-Senftenberg, Chair of Applied Physics and Semiconductor Spectroscopy, Konrad-Zuse-Str. 1, 03046 Cottbus, Germany*

* camilla.coletti@iit.it

## Abstract

The adoption of graphene in electronics, optoelectronics and photonics is hindered by the difficulty in obtaining high quality material on technologically-relevant substrates, over wafer-scale sizes and with metal contamination levels compatible with industrial requirements. To date, the direct growth of graphene on insulating substrates has proved to be challenging, usually requiring metal-catalysts or yielding defective graphene. In this work, we demonstrate a metal-free approach implemented in commercially available reactors to obtain high-quality monolayer graphene on c-plane sapphire substrates via chemical vapour deposition (CVD). We identify via low energy electron diffraction (LEED), low energy electron microscopy (LEEM) and scanning tunneling microscopy (STM) measurements the Al-rich reconstruction $(\sqrt{31} \times \sqrt{31})R \pm 9°$ of sapphire to be crucial for obtaining epitaxial graphene. Raman spectroscopy and electrical transport measurements reveal high-quality graphene with mobilities consistently above 2000 cm$^2$/Vs. We scale up the process to 4-inch and 6-inch wafer sizes and demonstrate that metal contamination levels are within the limits for back-end-of-line (BEOL) integration. The growth process introduced here establishes a method for the synthesis of wafer-scale graphene films on a technologically viable basis.

**Introduction**

The route for the implementation of graphene in the electronic/optoelectronic-technology market relies on the existence of a synthesis method that yields graphene films over wafer-scale, with good crystallinity and with contamination levels compatible with large-scale back-end-of-line (BEOL) integration. At present, chemical vapour deposition (CVD) on catalytic copper (Cu) substrates is widely recognized as the most promising route to obtain scalable monolayer graphene for electronic and optoelectronic applications[1–4]. However, significant hurdles are limiting the actual integration of CVD graphene grown on Cu for most applications. In the first instance, the unavoidable transfer process over wafer-scale is rather cumbersome and introduces contamination, unintentional doping, and mechanical stress[5-7], which adversely impact the physical integrity and electrical performance[8] of the graphene layer. The significant challenge involved in carrying out this seemingly straightforward task, is reflected by the vast literature on large-scale transfer processes. Secondly, metallic contamination levels in transferred CVD graphene grown on Cu are typically well-above the specifications requested for BEOL integration[6]. Clearly, as graphene moves closer towards applications, contamination will become an increasingly serious roadblock, unless addressed.

A way to overcome the above hurdles would be to directly synthesize graphene onto the target substrate, such as epitaxial graphene on silicon carbide (SiC)[9,10]. There, high growth temperatures provide a sufficient amount of energy to sublimate silicon from the substrate, while the remaining carbon rearranges on the surface in the form of graphene. However, the very high cost of the substrate and of the process itself, together with its marginal and niche application range in consumer electronics make it something of a *cul-de-sac* as a route for commercialisation.

The successful synthesis of monolayer high-quality graphene on sapphire would instead be readily implemented into what is already a very mature device-processing technology as well as into the vast market pushing it. Indeed, sapphire has recently become ubiquitous as a substrate for light-emitting diodes[11-13], and is being adopted in microelectronics with relevance in high frequency and data communication[13]. Also, synthesis of graphene on sapphire would provide an alternative route to obtain metal-free graphene that could be transferred onto final target substrates, something that to date has not been achieved at wafer scale for epitaxial graphene on SiC due to the very strong epitaxial interaction with the growth substrate. To date, several works have reported attempts to synthesize graphene directly onto insulating substrates, mostly silicon and sapphire[14-24]. Most of them use metal catalysts sacrificially deposited on the substrate[16, 20–23] or in the vapour phase[18] to aid the growth, which does not resolve the metallic contamination issue. Only few works have reported the metal-free synthesis of graphene on sapphire, obtaining high-quality graphene over small areas at growth temperatures higher than 1500 °C[14,15]. Scaling up high-quality metal-free graphene on sapphire has proved to be challenging, with best reported mobilities for 2 inch graphene wafers of about 370 cm$^2$/Vs[17]. Also, no work has identified to date the sapphire surface reconstruction upon graphene growth, a crucial aspect in identifying and clarifying favourable growth mechanisms. Ultimately, to date, no work has identified a clear path to obtain wafer-scale metal-free graphene on sapphire with mobilities comparable to those obtained for graphene grown on Cu. Here, we demonstrate and scale up to 4-inch and 6-inch wafers a CVD metal-free approach for growing graphene directly on sapphire substrates that yields films with mobilities above 2000 cm$^2$/Vs and contamination levels compatible with BEOL integration. We show that wafer-scale graphene films grown on sapphire can be transferred with a metal-free approach while maintaining the original (as-grown) carrier mobilities. Furthermore, we perform an in-depth investigation of the graphene/sapphire interface via low energy electron diffraction (LEED) and scanning tunneling microscopy (STM), which allows us to identify the path for high-quality epitaxial growth.

## Results

### The importance of surface preparation

Two different approaches for graphene synthesis were adopted and compared as shown in the schematic diagram reported in Fig 1 (a). In one case, graphene was grown directly on sapphire without preparing the surface (pristine), while in the other one the sapphire surface was hydrogen-etched (H$_2$-etched) before graphene deposition. In both cases, c-plane Al$_2$O$_3$(0001) dies were introduced in a high temperature cold-wall research reactor (AIXTRON BM Pro HT) and graphene was grown at 1200 °C, 25 mbar, in a mixture of 20:2:0.1 Ar:H$_2$:CH$_4$ for 30 minutes. In the H$_2$-etched approach, an additional step was performed prior to graphene growth, where the sapphire samples were etched in the same reactor at 1180°C, 750 mbar, in H$_2$ atmosphere for 5 minutes. More detailed information about the growth and etching processes can be found in the Experimental Section. Raman spectroscopy, atomic force microscopy (AFM), and electrical measurements were performed to investigate the quality of the graphene grown with the two different approaches (Fig.1(b)-(g)). The topographical difference between the graphene grown on pristine and H$_2$-etched sapphire is remarkable, as visible in the AFM micrographs reported in Fig. 1 (b) and (c), respectively. The graphene film grown on pristine sapphire shows a high density of ridges (see panel (b)), similar to those measured on graphene on SiC(000-1)[25]. Such ridges form as a consequence of the different thermal expansion coefficients of graphene and sapphire, as also observed in other works[14,15,17], and have a height of around 1-4 nm (Fig. S1(a) in Supplementary Information). In addition, scratches originating from the substrate polishing process are also visible, even after graphene growth. In contrast, the graphene grown on H$_2$-etched sapphire, exhibits a significantly reduced density of ridges. Also, well-defined atomic steps with heights that are integer multiples of the single unit cell height (i.e., 1.3 nm) become visible (panel(c) and Fig. S1(b)). Hence, similarly to SiC[26], optimized H$_2$-etching reveals atomic terraces on sapphire. Histograms obtained from representative Raman maps acquired over areas of 25x25 $\mu m^2$ are reported in panels (d)-(g) and consistently indicate a significant improvement of the crystalline quality of graphene on $H_2$-etched sapphire (see also Fig. S2). Each Raman histogram is fitted with a Gaussian curve to extract the distribution maximum and the half-width-at-half-maximum, the latter employed as uncertainty. Fig. 1(d) shows that the full-width-at-half-maximum (FWHM) of the 2D mode of graphene decreases from 38.44 (±3.17) cm$^{-1}$ to 32.32 (±2.22) cm$^{-1}$ for graphene grown on pristine and $H_2$-etched sapphire, respectively, indicating an improvement in graphene crystalline quality and less strain fluctuation across the samples[27]. Notably, the average 2D FWHM measured on $H_2$-etched sapphire is the lowest reported to date for as-grown graphene on sapphire. The D/G intensity ratio, indicative of the defect concentration in graphene, presents a bimodal distribution with the main peak at 0.94 and a broader peak at 1.4, suggesting the presence of highly defective areas across the sample. Upon $H_2$-etching of the substrate, the D/G distribution peaks at a much-reduced value of 0.13 (±0.04) (Fig. 1(e)), indicating a low defect density. The 2D/G intensity ratio distributions peak at 1.80 (±0.17) and 3.67 (±0.27) for graphene on pristine and $H_2$-etched sapphire, respectively (Figure 2(f)). These values indicate a lower charge carrier concentration for graphene on the $H_2$-etched sapphire sample, which we estimate[28] being in the lower $10^{12}$ cm$^{-2}$ range, compared to graphene grown on pristine sapphire, which is estimated to be around 5x $10^{12}$ cm$^{-2}$. These estimates are confirmed by Hall effect measurements at room temperature. The highest carrier mobility measured for graphene on $H_2$-etched sapphire is 2260 cm$^2$/Vs, with a hole density of 2.3 x $10^{12}$ cm$^{-2}$. For graphene on pristine sapphire, the highest mobility is 890 cm$^2$/Vs, with a hole density of 5.24 x $10^{12}$ cm$^{-2}$. In general, we observe that mobilities on $H_2$-etched substrates are at least a factor of 2.5 higher than on non-treated substrates grown in nominally-identical conditions (see Supplementary Information for further details). It is therefore clear that $H_2$-etching of sapphire is crucial to obtain high-quality graphene while growing at temperatures comparable to those conventionally used for metal-CVD processes.

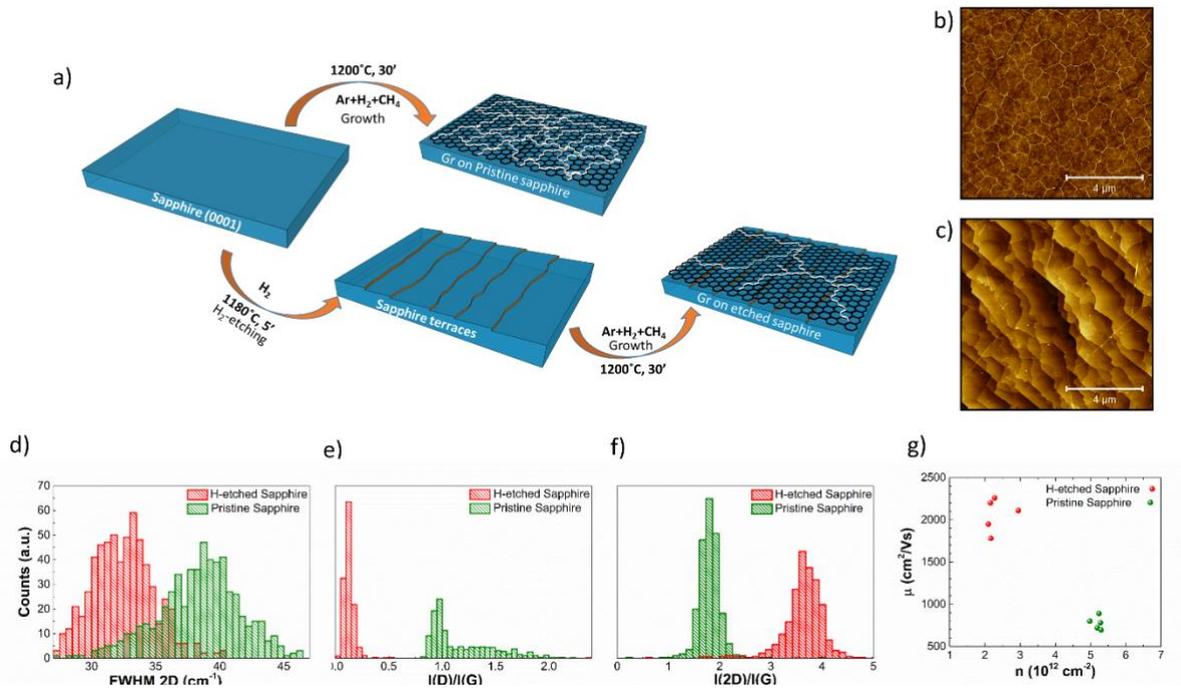

Figure 1. (a) Schematic sketch depicting the processing steps for the two different approaches for obtaining graphene: growth on pristine and $H_2$-etched sapphire. (b-c) Representative 10x10 μm AFM micrographs of graphene grown on (b) pristine and (c) $H_2$-etched sapphire, respectively. (d-f) Representative histograms obtained from 25x25 μm$^2$ Raman maps of graphene on pristine and $H_2$-etched sapphire showing the (d) FWHM of the 2D peak, (e) the D/G intensity ratio and (f) the 2D/G intensity ratio. (g) Mobility-vs-carrier density plot for graphene Hall bars fabricated on pristine and $H_2$-etched sapphire.

**Structural properties of the graphene/sapphire interface**

In order to determine the structural properties at the graphene/sapphire interface and gain a better understanding of the growth mechanisms – something that to date has remained elusive – we performed LEED and STM measurements. First, LEED was carried out on pristine and etched sapphire surfaces. For pristine sapphire surfaces, as expected, no LEED pattern could be retrieved, due to the insulating nature of the sample. On the other hand, surprisingly, LEED on etched sapphire was measurable down to 60 eV and revealed a clear $(\sqrt{31}\times\sqrt{31})R9°$ reconstruction, (see Fig. S4(a) in Supplementary Information). The Al-rich nature of tis reconstruction accounts for the direct visualization of the LEED pattern. Although there is a rather long history of controversy and debates concerning its actual atomic structure [29–33], Lauritsen et al. bring a compelling argument in favour of an Al(111) layer on top of the Al-terminated c-plane of the substrate[30]. To date, this reconstruction has been observed only upon annealing in ultra-high vacuum (UHV) of $Al_2O_3(0001)$ at temperatures well-above 1200°C[29–32], resulting in a loss of surface oxygen. Our recipe can induce the $(\sqrt{31}\times\sqrt{31})R9°$ reconstruction at lower temperatures and higher pressures than those reported in literature[29–32] as a result of the $H_2$-etching process, i.e., oxygen reduction is facilitated by hydrogen. Figure 2(a) reports the LEED pattern recorded after graphene synthesis on etched sapphire and, as visible in Fig. 2(b), all measured spot positions are perfectly explained by the expected presence of two rotational domains of the (√31×√31)R9° reconstruction (i.e., (√31×√31)R±9°). In Fig. 2 (c) we show an atomically-resolved STM image of graphene over the (√31×√31)R±9° reconstruction. The complex pattern of the underlying reconstructed sapphire layer is recognizable as a periodic arrangement of irregularly rhombi-shaped domains. In the bottom inset of the panel, we show the line profile (red solid line on the image) indicating the spacing between the maxima to be (27±1) Å, compatible with the nominal length of the reconstruction of 26.5 Å. A 2D-FFT filtered portion of the image in panel (c) is shown in panel (d) together with the directions of the graphene (red) and the reconstruction (white) primitive vectors. The mutual orientation between the two is 21°, corroborating the fact that graphene preferentially aligns along the R30 direction with respect to the $Al_2O_3$(0001) (1×1), as also visible from the LEED pattern in panel (a). The structural model of the Al-rich reconstruction on sapphire is complicated by the fact that the lattice spacing between the Al atoms is not constant within the unit cell, but depends on the mutual arrangement of the Al atoms with respect to the substrate registry[30]. Although the current resolution of the STM measurements does not allow us to draw a conclusion on the atomic arrangement between the graphene and the Al atoms at the interface, we observe that the graphene lattice conforms well to the periodic corrugation imposed by the (√31×√31)R±9°. Thus, graphene growth on sapphire is apparently catalysed by the highly reactive Al-rich surface of the reconstructed surface. During growth, the Al sites are strong Lewis acids that dissociate the methane molecules, thus catalyzing graphene growth.

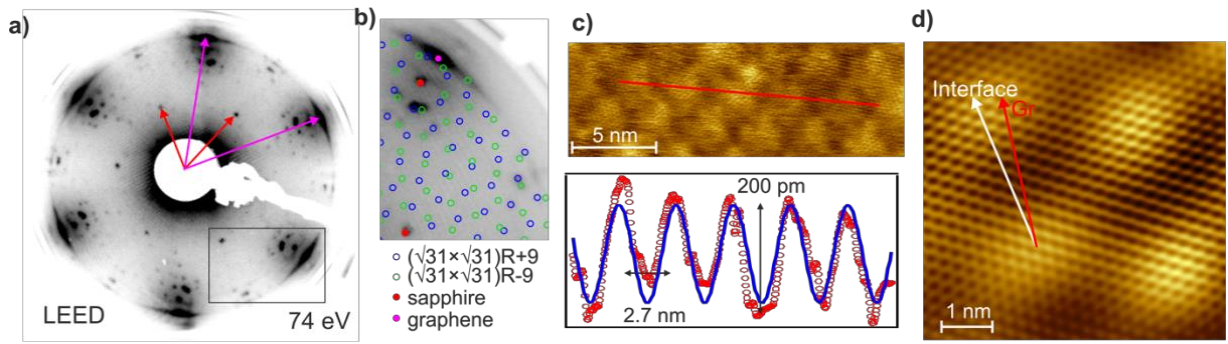

Figure 2: (a) LEED pattern of graphene grown on $H_2$-etched sapphire, measured at 74 eV. (b) Zoom of the inset in panel (a), showing the sapphire reconstruction superimposed with the theoretical diffraction spots. (c) STM image and a line profile (along the red solid line) of a portion of graphene over the (√31×√31)R±9°. (d) FFT-Filtered STM image from a portion of (c).

To assess the size of the single-crystalline graphene domains, the sample was measured with low-energy electron microscopy (LEEM), which is a technique highly sensitive to the crystalline orientation[34]. In the LEEM micrograph shown in Fig. S5, two phases (light and dark gray) are mainly visible. When performing low-energy electron diffraction with micro-spot illumination (µLEED) on the light gray phase, only the (√31×√31)R9° pattern is recognizable, together with the R30 graphene reflections. On the dark gray phase instead, multiple rotational domains are found, and no (√31×√31)R9° pattern is observed. From the dark-field analysis reported in the Supplementary Information (see Fig. S5(b)), we estimate the single-crystal grain size of graphene to be of the order of one micrometer. This is about one order of magnitude larger than what has been reported in the literature so far [15,17] and we argue that it is related to the micron-sized domains of the Al-rich reconstructed surface. Hence, on one hand graphene grows with a high degree of crystallinity on the fully reconstructed sapphire surface and, on the other hand, the graphene single-crystal domain size is essentially limited by the grain size of the reconstructed domains on the $Al_2O_3$(0001) surface. Therefore, the route for obtaining high quality graphene on sapphire relies on the fine control of the Al-rich reconstructed sapphire surface.

**Scalability up to 6-inch wafers**

To demonstrate the industrial viability of this method, the process is transferred onto a production-scale reactor (AIXTRON CCS 2D) and graphene growth on sapphire is demonstrated in 5 x 4-inch and 1 x 6-inch configuration. Figure 3 shows the results of the Raman analysis of graphene grown on a 6-inch sapphire wafer. The analysis is carried out on nine areas of 2 x 2 $mm^2$, as indicated by the squares superposed to the optical picture of the as-grown wafer in Figure 3(a). Figure 3(b) shows the histograms of the FWHM of the graphene 2D Raman peak and of the ratio of the intensities of the D and G peaks for five selected areas (center and even-numbered quadrants). These Raman histograms are employed to benchmark the graphene crystalline quality on the different areas of the wafer, in terms of concentration of defects and strain-doping fluctuation. The comparison of the different histograms of the 2D FWHM demonstrates a high degree of homogeneity of the graphene film throughout the wafer. The average value of the 2D FWHM is 36 $cm^{-1}$ for all the areas. Remarkably, the D/G peak intensity ratio average ranges from 0.15 ± 0.06 to 0.2 ± 0.12, demonstrating that even the defect concentration shows little variation across the wafer. The 2D/G peak intensity ratio average varies between 3.45 ± 0.30 and 3.83 ± 0.35 (see Supplementary Information), confirming that the synthesized material is monolayer. The complete histograms and the precise values of the 2D FWHM, D/G and the 2D/G peak intensity ratios for the 6-inch and 4-inch wafers are reported in the Supplementary Information. In all cases, the wafers processed at the same time showed the same quality and homogeneity, and the process repeatability was outstanding. In all the samples analysed, the (√31×√31)R±9° reconstruction could be observed with LEED.

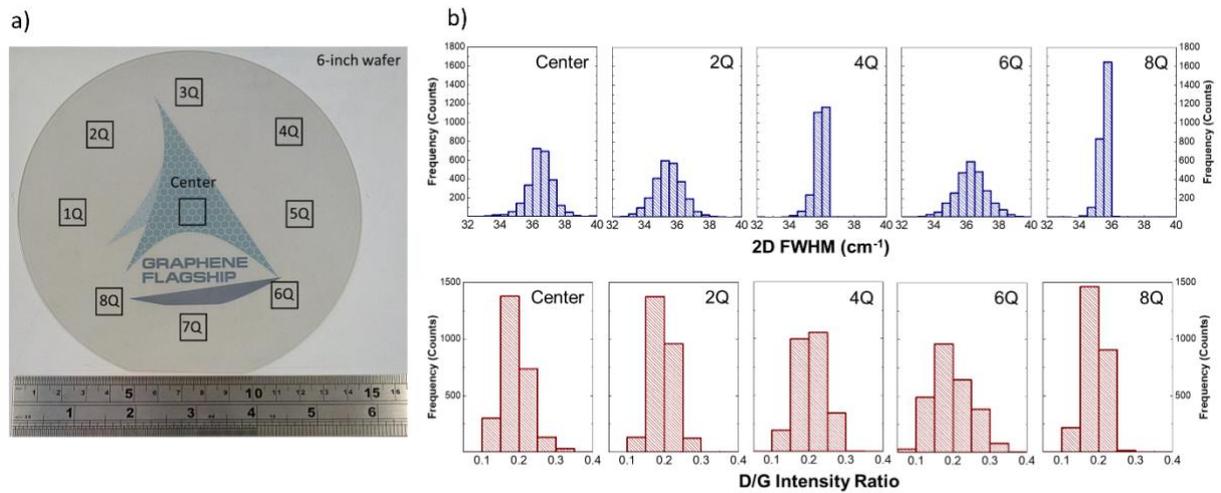

Figure 3. Characterization of graphene grown on 6-inch sapphire wafer. (a) Optical image of the grown wafer, with squares indicating the approximate locations where Raman measurements were performed. Sample looks highly transparent, with the "Graphene Flagship" logo behind remaining clearly visible. (b) Raman analysis of graphene grown on a 6-inch sapphire wafer, measured at 5 different places.

To measure the conductivity of the graphene on sapphire wafers, we performed terahertz time-domain spectroscopy (THz-TDS) measurements[35] (see Experimental Section for additional information). Fig. 4(a) shows the sheet conductivity map of an entire 4-inch wafer, and the average sheet conductivity value, retrieved by fitting the histogram in Fig. 4(b), is 0.91±0.04 mS. To broaden the applicative range, graphene grown on sapphire wafers was successfully transferred to target substrates (in this case 90 nm $SiO_2$/Si) using the PVA lamination approach[36]. Fig. 4(c) shows a 4-inch graphene film transferred onto $SiO_2$/Si, continuous over 97% of the area. Field effect transistors (FET) were fabricated from graphene transferred on $SiO_2$/Si dies and measured in a backgate geometry (Fig. 4(e)). Fig. 4(d) shows a typical measured device, consisting in a series of stripes of graphene with lateral size of 3.7 μm and lengths variable between 33 μm and 42 μm, contacted with 10/60 nm of Cr/Au contacts. A typical conductivity versus gate voltage curve is reported in Fig. 4(f): the neutrality point is around -5 V, indicating low doping of graphene (n ≈ 3 x $10^{11}$ $cm^{-2}$). Mobility values are obtained from the slope of the linear fit of the conductivity versus gate voltage[37], according to the formula; $\mu = \pm \frac{\sigma t}{\epsilon \epsilon_0 (V_g - V_D)}$, where t and ε are the thickness and the dielectric constant of $SiO_2$, respectively. For the representative device shown in Fig. 4(d), we obtain $\mu_{h,}$ = 2300 $cm^2$/Vs for holes and $\mu_e$ = 2000 $cm^2$/Vs for electrons, respectively (see also Figs. S8-10). These values are comparable with what has been reported for CVD polycrystalline graphene grown on Cu foil and transferred onto $SiO_2$/Si[38]. However, in contrast to CVD graphene on Cu, these samples already fully satisfy the specifications for BEOL integration as confirmed by total reflection X-ray fluorescence (TXRF) measurements (see Supplementary Information), (even if manually handled under non-controlled laboratory conditions).

We are confident that even front-end-of-line (FEOL) requirements could be met with more stringent wafer preparation and handling under fab conditions, since there is no metal used for the growth or transfer processes in our method.

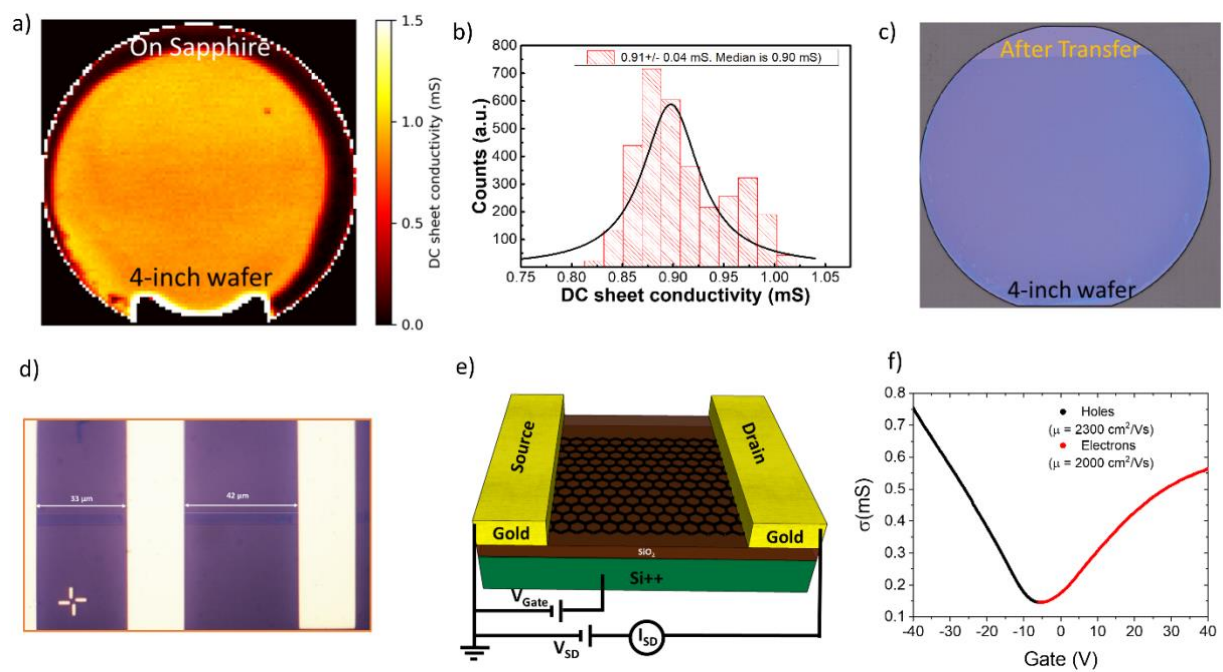

Figure 4. (a) THz-TDS mapping of the sheet conductivity across the 4-inch wafer. (b) Histogram of THz-TDS sheet conductivity measured of graphene on sapphire. (c) Optical image of graphene transferred onto a 90nm-SiO$_2$/Si substrate by the PVA lamination method. (d) Optical view of the fabricated devices for electrical characterization of graphene grown on sapphire and transferred to SiO$_2$/Si. The yellow stripes are the metal contacts and the darker horizontal areas are the two graphene stripes defined by EBL and RIE processing (marked up by white lines). (e) Schematic diagram showing the FET device and the electrical setup. A highly doped silicon substrate, covered with 285 nm of dielectric oxide, is used as a backgate for the FET characterization. (f) Two-probe conductivity σ as a function of the applied backgate voltage for electrons and holes (in red and black respectively).

## Discussion

In conclusion, this work identifies a clear approach for obtaining graphene directly on the c-plane of Al$_2$O$_3$(0001) substrates in commercially available CVD reactors. We show that no external catalyst needs to be added in the growth process to obtain high-quality graphene, if the sapphire surface is properly prepared. Indeed, sapphire preparation via H$_2$-etching is crucial to obtain an Al-rich ($\sqrt{31}$ x $\sqrt{31}$)R± 9° reconstructed surface, which catalyzes graphene growth at temperatures comparable to those conventionally used for metal-CVD processes (i.e. 1200 °C). We show for the first-time that high-quality graphene can be grown on on 4-inch and 6-inch sapphire wafers with quality and properties comparable to those obtained for graphene on Cu foil. The results show a high degree of uniformity and consistency, which is crucial for any industrial process. While we here demonstrate up to 6-inch wafer growth, there is no indication that commercially available 12'' sapphire substrates should not produce similar results. The clear advantage of this approach is the compatibility with fab contamination specifications, as shown by TXRF measurements, and the straightforward use of readily available sapphire wafer substrates (in contrast to metal foils, or thin films on wafer, or SiC). Furthermore, we demonstrate that large wafer areas of graphene can be transferred to any target substrate with a polymeric lamination approach, and that the obtained carrier mobilities are about 2000 cm$^2$/Vs. LEEM measurements indicate that the synthesized graphene is polycrystalline, with a preferential orientation of 30° with respect to the Al$_2$O$_3$(0001) substrate and with grain sizes in the micrometer range, which is about one order of magnitude larger than previously reported. Having observed to what extent graphene is affected by the ordering and composition of the interface, we suggest that a fine control over the homogeneity and domain size of the reconstructed sapphire surface should enhance the quality as well as the domain size of the grown graphene even further, thereby increasing its electrical mobility. This work demonstrates a viable route for the production of metal-free wafer-scale high-quality graphene directly onto insulating substrates, with a significant potential impact on a wide number of microelectronic, optoelectronic, and photonic applications.

**Methods**

**Growth of Gr on sapphire**

We used c-axis, HEMCOR single crystal, double side polished sapphire (0001) substrates supplied by Alfa Aesar, (Germany). Before growth, sapphire substrates were cleaned with acetone, isopropanol, and de-ionized (DI) water in an ultra-sonicator bath, immersed in piranha solution (1:3, H$_2$O$_2$:H$_2$SO$_4$) for 15 min and finally washed in DI water and N$_2$-blow-dried. The H$_2$-etching was performed in the growth reactor at 1180°C for 5 min in an atmosphere of H$_2$[39]. Samples were then extracted and characterized. Graphene growth for both H$_2$-etched and pristine sapphire was performed as follows: (i) the substrate was annealed at 1200°C for 10 min in an atmosphere of 1000 sccm of Ar at 25 mbar; (ii) growth was performed by introducing 100 sccm H$_2$ and 5 sccm of CH$_4$ while flowing 1000 sccm of Ar for 30 min at 25 mbar; (iii) cooling was carried out under Ar flux.

**Characterization**

Raman measurements were performed with a Renishaw Invia system with a 532 nm laser, a spot size of ~1 μm, and at 5 mW of laser power. AFM was carried out with an AFM+ (Bruker Dimension Icon) operated in tapping mode in air and the Gwyddion software package was used to analyse the micrographs. LEED measurements were performed at room temperature with a SPECS GmbH LEED optics. STM and scanning tunneling spectroscopy (STS) measurements were carried out in an Omicron LT-STM at a base pressure of 10$^{-10}$ mbar. LEEM measurements were done using an Elmitec LEEM III microscope with energy filter operated at an electron energy of 15 keV and a base pressure of 10$^{-10}$ mbar. In μLEED, the incident electron beam was limited to an area of about 250 nm in diameter.

The electrical transport measurements on the transferred graphene were performed in a home-made probing station on an optical table to minimize vibrations. The gate and drain voltages were provided by a couple of Keithley 2450, used also to measure the source-drain current and the eventual presence of leak current between the gate and the drain. The device was contacted using tungsten tips of 25µm radius, aligned using 3 MPI MP-40 micropositioner. Field effect transistors (FET) were defined by electron beam lithography (EBL) and reactive ion etching (RIE) techniques. All the electrical characterizations were performed in air at room temperature, in a 2-probe configuration applying a DC bias between source and drain of 10 mV. Before measuring, the samples were annealed in UHV at 230 °C for 2 hours.

Terahertz time-domain spectroscopy (THz-TDS) of as-grown graphene on sapphire was conducted in transmission mode using a commercial Picometrix T-Ray 4000 system with a THz spot size of ~350 µm at 1 THz[35]. Samples were raster scanned with 1 mm step size in the focal plane between the THz transmitter and receiver. Examples of THz time-domain waveforms are shown in Figure S9(a). The waveforms contain transients from internal reflections within the sapphire substrate. Here, we use the data from the directly transmitted transients to extract the sheet conductivity ($\sigma_s(\omega) = \sigma_1 + i\sigma_2$) of graphene as[40–42]

$$\sigma_s(\omega) = \left(\frac{1}{Z_0 T_{meas}(\omega)} - 1\right)(n_{sap} + 1),$$

where $Z_0$ is the vacuum impedance, $T_{meas}$ is ratio of the Fourier transforms of the THz waveforms transmitted through graphene-covered sapphire and bare sapphire (Figure S9(b)), and $n_{sap}$ is the refractive index of sapphire (Figure S9(c)) calculated from THz waveforms from bare sapphire relative to air[43].

Examples of sheet conductivity spectra from as-grown graphene on sapphire are shown in Supporting Figure S9(d). The sheet conductivity spectra do not follow the classical Drude-model as previously observed[41,42,44] as we notice a reduction in the sheet conductivity at low frequencies – this indicates that the carriers in graphene do not scatter isotropically but experience some degree of carrier localization[40,42,45]. In such cases, the sheet conductivity is better described by the first term of the phenomenological Drude-Smith model[40,42,46]

$$\sigma_s(\omega) = \frac{W_D}{(1 - i\omega\tau)}\left(1 + \frac{c}{(1 - i\omega\tau)}\right),$$

where $W_D$ is the Drude weight related to the DC sheet conductivity as $\sigma_{DC} = W_D(1 + c)$, and $c$ is a parameter that can take values from -1 to 0 and describes the degree of carrier localization/backscattering[40,42]. If $c = 0$, the carrier momentum is totally randomized (classical Drude model), while carriers are completely backscattered in the case $c = -1$[45]. Fits to the Drude-Smith model and the extracted parameters for $\sigma_{DC}$, $\tau$, and $c$ are shown together with the sheet conductivity spectra in Supporting Figure S9(d).

**Acknowledgements**

The research leading to these results has received funding from the European Union′s Horizon 2020 research and innovation program under grant agreements Nos. 696656 – GrapheneCore1 and 785219 – GrapheneCore2. A.S. and P.B. acknowledge support from the Danish National Research Foundation Center of Excellence for Nanostructured Graphene (CNG) (project DNRF103).

**Author contributions**

N.M, S.F., F.F., L.M, C.M., B.C, P.R.W., A.S, L.B, I.A., S.R. performed the experiments. J.F. and J.I.F. supervised LEEM measurements. P.B., K.T and C.C designed and supervised the study. C.C. coordinated the project. All the authors discussed the data and wrote the paper.

**References**

1. Bae, S. *et al.* Roll-to-roll production of 30-inch graphene films for transparent electrodes. *Nat. Nanotechnol.* **5**, 574–578 (2010).

2. Rahimi, S. *et al.* Toward 300 mm wafer-scalable high-performance polycrystalline chemical vapor deposited graphene transistors. *ACS Nano* **8**, 10471–10479 (2014).

3. Miseikis, V. *et al.* Deterministic patterned growth of high-mobility large-crystal graphene: a path towards wafer scale integration. *2D Mater.* **4**, 021004 (2017).

4. Romagnoli, M. *et al.* Graphene-based integrated photonics for next-generation datacom and telecom. *Nat. Rev. Mater.* **3**, 392–414 (2018).

5. Bonaccorso, F. *et al.* Production and processing of graphene and 2d crystals. *Mater. Today* **15**, 564–589 (2012).

6. Lupina, G. *et al.* Residual Metallic Contamination of Transferred Chemical Vapor Deposited Graphene. *ACS Nano* **9**, 4776–4785 (2015).

7. Chen, Y., Gong, X.-L. & Gai, J.-G. Progress and Challenges in Transfer of Large-Area Graphene Films. *Adv. Sci.* **3**, 1500343 (2016).

8. Bruna, M. *et al.* Doping Dependence of the Raman Spectrum of Defected Graphene. *ACS Nano* **8**, 7432–7441 (2014).

9. Berger, C. *et al.* Ultrathin Epitaxial Graphite: 2D Electron Gas Properties and a Route toward Graphene-based Nanoelectronics. *J. Phys. Chem. B* **108**, 19912–19916 (2004).

10. Emtsev, K. V. *et al.* Towards wafer-size graphene layers by atmospheric pressure graphitization of silicon carbide. *Nat. Mater.* **8**, 203–207 (2009).

11. Amano, H., Asahi, T. & Akasaki, I. Stimulated Emission Near Ultraviolet at Room Temperature from a GaN Film Grown on Sapphire by MOVPE Using an AlN Buffer Layer. *Jpn. J. Appl. Phys.* **29**, L205–L206 (1990).

12. Chen, Z. *et al.* High-Brightness Blue Light-Emitting Diodes Enabled by a Directly Grown Graphene Buffer Layer. *Adv. Mater.* **30**, 1801608 (2018).

13. https://www.reportsmonitor.com/report/420029/LED-Sapphire-Substrate-Market

14. Fanton, M. A. *et al.* Characterization of Graphene Films and Transistors Grown on Sapphire by Metal-Free Chemical Vapor Deposition. *ACS Nano* **5**, 8062–8069 (2011).

15. Hwang, J. *et al.* van der Waals Epitaxial Growth of Graphene on Sapphire by Chemical Vapor Deposition without a Metal Catalyst. *ACS Nano* **7**, 385–395 (2013).

16. Ago, H. *et al.* Epitaxial Chemical Vapor Deposition Growth of Single-Layer Graphene over Cobalt Film Crystallized on Sapphire. *ACS Nano* **4**, 7407–7414 (2010).

17. Song, H. J. *et al.* Large scale metal-free synthesis of graphene on sapphire and transfer-free device fabrication. *Nanoscale* **4**, 3050 (2012).

18. Kim, H. *et al.* Copper-Vapor-Assisted Chemical Vapor Deposition for High-Quality and Metal-Free Single-Layer Graphene on Amorphous SiO 2 Substrate. *ACS Nano* **7**, 6575–6582 (2013).

19. Chen, J. *et al.* Oxygen-aided synthesis of polycrystalline graphene on silicon dioxide substrates. *J. Am. Chem. Soc.* **133**, 17548–17551 (2011).

20. Chen, Y., Gong, X.-L. & Gai, J.-G. Progress and Challenges in Transfer of Large-Area Graphene Films. *Adv. Sci.* **3**, 1500343 (2016).

21. Ning, J. *et al.* Review on mechanism of directly fabricating wafer-scale graphene on dielectric substrates by chemical vapor deposition. *Nanotechnology* **28**, 284001 (2017).

22. Zhang, C. *et al.* Transfer-free growth of graphene on $Al_2O_3$ (0001) using a three-step method. *Carbon N. Y.* **131**, 10–17 (2018).

23. Kim, Y. *et al.* Synthesis of high quality graphene on capped (1 1 1) Cu thin films obtained by high temperature secondary grain growth on c -plane sapphire substrates. *2D Mater.* **5**, 035008 (2018).

24. Muñoz, R. *et al.* Direct synthesis of graphene on silicon oxide by low temperature plasma enhanced chemical vapor deposition. *Nanoscale* **10**, 12779–12787 (2018).

25. de Heer, W. A. *et al.* Large area and structured epitaxial graphene produced by confinement controlled sublimation of silicon carbide. *Proc. Natl. Acad. Sci.* **108**, 16900–16905 (2011).

26. Frewin, C. L., Coletti, C., Riedl, C., Starke, U. & Saddow, S. E. A Comprehensive Study of Hydrogen Etching on the Major SiC Polytypes and Crystal Orientations. *Mater. Sci. Forum* **615–617**, 589–592 (2009).

27. Ferrari, A. C. & Basko, D. M. Raman spectroscopy as a versatile tool for studying the properties of graphene. *Nat. Nanotechnol.* **8**, 235–246 (2013).

28. Das, A. *et al.* Monitoring dopants by Raman scattering in an electrochemically top-gated graphene transistor. *Nat. Nanotechnol.* **3**, 210–215 (2008).

29. Gautier, M., Duraud, J. P., Pham Van, L. & Guittet, M. J. Modifications of α-Al2O3(0001) surfaces induced by thermal treatments or ion bombardment. *Surf. Sci.* 250, 71–80 (1991).

30. Lauritsen, J. V. *et al.* Atomic-Scale Structure and Stability of the Atomic-Scale Structure and Stability of the $\sqrt{31} \times \sqrt{31}$ R9° Surface of Al2O3(0001) *Phys. Rev. Lett.* **103**, 076103 (2009).

31. Renaud, G., et al. Atomic Structure of the α−Al2O3(0001) ($\sqrt{31} \times \sqrt{31}$)R±9°Reconstruction *Phys. Rev. Lett.* **73**, 1825–1828 (1994).

32. Barth, C. & Reichling, M. Imaging the atomic arrangements on the high-temperature reconstructed α-Al2O3(0001) surface. *Nature* **414**, 54–57 (2001).

33. Jarvis, E. A. A. & Carter, E. A. Metallic Character of the Al 2 O 3 (0001)-($\sqrt{31} \times \sqrt{31}$) R ± 9° Surface Reconstruction †. *J. Phys. Chem. B* **105**, 4045–4052 (2001).

34. Bauer, E., Mundschau, M., Swiech, W. & Telieps, W. Surface studies by low-energy electron microscopy (LEEM) and conventional UV photoemission electron microscopy (PEEM). *Ultramicroscopy* **31**, 49–57 (1989).

35. Buron, J. D. *et al.* Graphene Conductance Uniformity Mapping. *Nano Lett.* **12**, 5074–5081 (2012).

36. Shivayogimath, A. *et al.* Do-It-Yourself Transfer of Large-Area Graphene Using an Office Laminator and Water. *Chem. Mater.* **31**, 2328–2336 (2019).

37. Hwang, E. H., Adam, S. & Sarma, S. Das. Carrier Transport in Two-Dimensional Graphene Layers. *Phys. Rev. Lett.* **98**, 186806 (2007).

38. Tsen, A. W. *et al.* Tailoring Electrical Transport Across Grain Boundaries in Polycrystalline Graphene. *Science.* **336**, 1143–1146 (2012).

39. Mishra, N. *et al.* Rapid and catalyst-free van der Waals epitaxy of graphene on hexagonal boron nitride. *Carbon.* **96**, 497–502 (2016).

40. Buron, J. D. *et al.* Terahertz wafer-scale mobility mapping of graphene on insulating substrates without a gate. *Opt. Express* **23**, 30721 (2015).

41. Whelan, P. R. *et al.* Electrical Homogeneity Mapping of Epitaxial Graphene on Silicon Carbide. *ACS Appl. Mater. Interfaces* **10**, 31641–31647 (2018).

42. Bøggild, P. *et al.* Mapping the electrical properties of large-area graphene. *2D Mater.* **4**, 042003 (2017).

43. Jepsen, P. U. & Fischer, B. M. Dynamic range in terahertz time-domain transmission and reflection spectroscopy. *Opt. Lett.* **30**, 29 (2005).

44. Whelan, P. R. *et al.* Robust mapping of electrical properties of graphene from terahertz time-domain spectroscopy with timing jitter correction. *Opt. Express* **25**, 2725 (2017).

45. Dadrasnia, E., Lamela, H., Kuppam, M. B., Garet, F. & Coutaz, J.-L. Determination of the DC Electrical Conductivity of Multiwalled Carbon Nanotube Films and Graphene Layers from Noncontact Time-Domain Terahertz Measurements. *Adv. Condens. Matter Phys.* **2014**, 1–6 (2014).

46. Smith, N. Classical generalization of the Drude formula for the optical conductivity. *Phys. Rev. B* **64**, 155106 (2001).

# Supporting Information

## Going beyond copper: wafer-scale synthesis of graphene on sapphire

N. Mishra[1,2], S. Forti[1], F. Fabbri[1,2], L. Martini[1], C. McAleese[3], B. Conran[3], P.R. Whelan[4,5], A. Shivayogimath[4,5], L. Buß[6], J. Falta[6], I. Aliaj[7], S. Roddaro[7,8], J. I. Flege[6,9], P. Bøggild[4,5], K.B.K. Teo[3] and C. Coletti[1,2,*]

[1]*Center for Nanotechnology Innovation @ NEST, Istituto Italiano di Tecnologia, Piazza San Silvestro 12, 56127 Pisa, Italy*
[2]*Graphene Labs, Istituto Italiano di Tecnologia, Via Morego 30, 16163 Genova, Italy*
[3]*AIXTRON Ltd., Buckingway Business Park, Anderson Rd, Swavesey, Cambridge CB24 4FQ, UK*
[4]*DTU Physics, Ørsteds Plads 345C, 2800 Kgs. Lyngby, Denmark*
[5]*Center for Nanostructured Graphene (CNG), Ørsteds Plads 345C, 2800 Kgs. Lyngby, Denmark*
[6]*Institute of Solid State Physics, University of Bremen, Bremen-28334, Germany*
[7]*NEST, Scuola Normale Superiore and Istituto Nanoscienze-CNR, Piazza S. Silvestro 12, I-56127 Pisa, Italy*
[8]*Dipartimento di Fisica, Università di Pisa, Largo B. Pontecorvo 3, 56127, Pisa (PI), Italy*
[9]*Brandenburg University of Technology Cottbus-Senftenberg, Chair of Applied Physics and Semiconductor Spectroscopy, Konrad-Zuse-Str. 1, 03046 Cottbus, Germany*

**AFM and Raman analysis of graphene on pristine and etched sapphire**

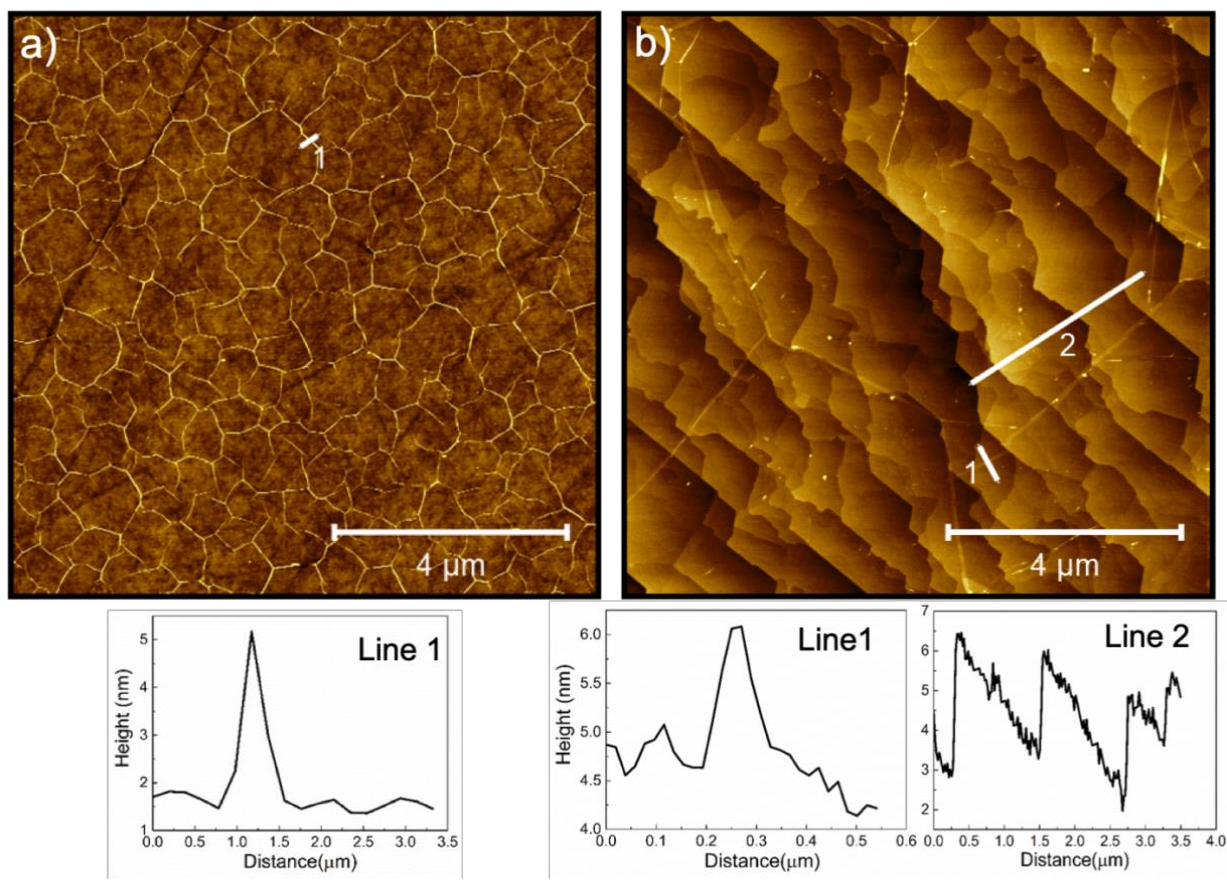

Figure S1. AFM line profile of wrinkles and terraces for graphene on (a) pristine and (b) $H_2$-etched sapphire surface for Figures 1(b) and 1(c).

Figure S1 (a) shows the height of graphene wrinkles formed due to different thermal expansion coefficients of sapphire and graphene. The typical height of those wrinkles is about 1-4 nm. Figure S1(b) shows the reduced wrinkling observed on graphene grown on $H_2$-etched sapphire and the atomic steps of sapphire revealed by the etching process. Line profile inserts reveal that the typical height of these terraces is about 1.5-4 nm. This is consistent with the fact that the reconstructed Al layer on the sapphire c-plane has an areal atomic density between 2.5 and 3 times higher than the sapphire basal plane[1], corresponding to unit cell decomposition between 1.25 and 1.5. The height of the wrinkles is again about 1-4 nm, as measured in Figure S1(b). In Figure S2 representative Raman spectra for the two types of samples are reported.

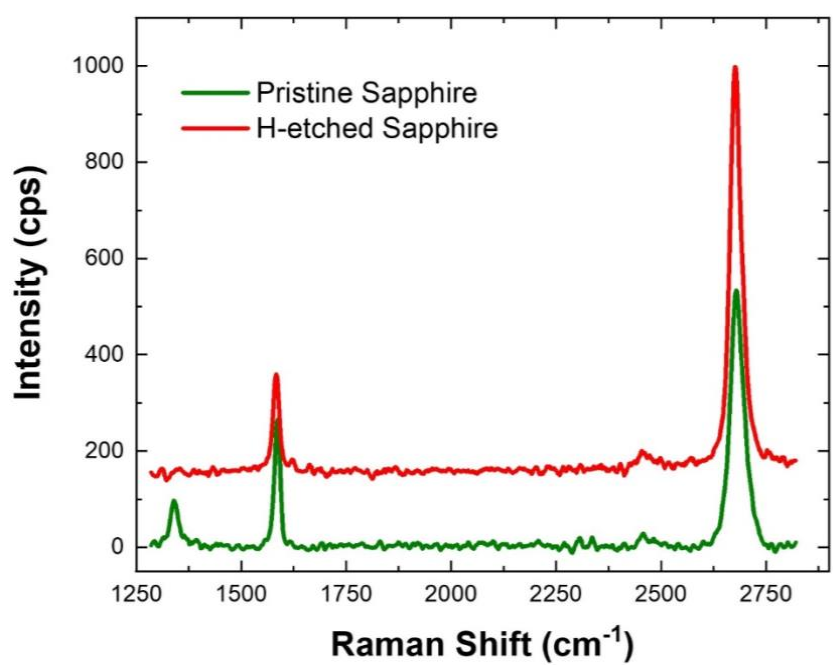

Figure S2. Representative Raman spectra of graphene on H-etched (red) and pristine (green) sapphire.

**Hall effect characterization of Gr/sapphire:**

The transport parameters of graphene films were measured at different locations in the chip using magneto-transport techniques. Several Hall bar devices were defined on the same chip by means of electron beam lithography (EBL), etching and metal deposition (10/50 nm Cr/Au) (Figure S2). The devices were electrically characterized at room temperature, in air and in dark conditions, using a lock-in amplifier operated at low frequencies (17 Hz). In particular, a 100 nA AC current was driven between a pair of contacts (source-drain) and the resulting voltage drops in the longitudinal and transverse direction were measured as a function of magnetic field in the range -0.37 to 0.37 T. The carrier densities and mobilities in each device were calculated from the magnetic field dependence of the transverse resistance and from the longitudinal resistance, respectively.

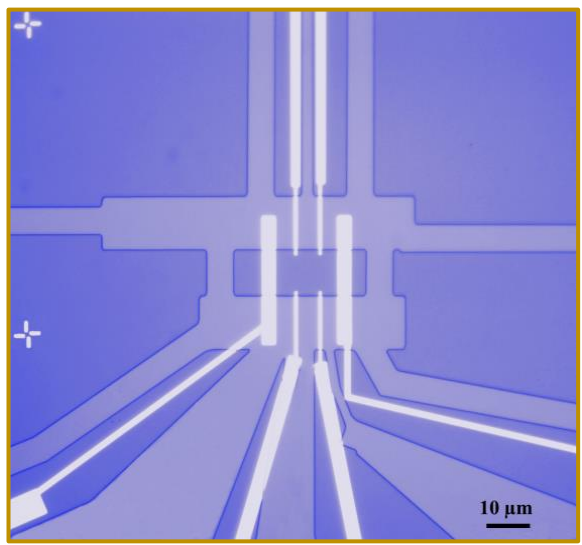

Figure S3. Hall Bar fabrication for transport measurements. The bright lines are Cr/Au contacts and the shallow purple areas are the etch mask that defines the final geometry of the graphene device.

Mobility and carrier density values plotted in Figure 1(g) are reported in Table S1. The general trend is that mobility on $H_2$-etched sapphire substrates is higher by a factor of about 2.5 with respect to that measured on pristine sapphire. Carrier density is also reduced for graphene on $H_2$-etched sapphire.

| Pristine | | $H_2$-etched | |
|---|---|---|---|
| n (cm$^{-2}$) *10$^{12}$ | µ(cm$^2$/Vs) | n (cm$^{-2}$)*10$^{12}$ | µ(cm$^2$/Vs) |
| 4.98 | 800 | 2.28 | 2260 |
| 5.28 | 780 | 2.16 | 2200 |
| 5.3 | 695 | 2.95 | 2110 |
| 5.24 | 890 | 2.17 | 1780 |
| 5.19 | 720 | 2.1 | 1950 |

Table S1. Carrier mobility and density for different Hall bars fabricated on graphene on pristine and $H_2$-etched sapphire.

# LEED analysis

Figure S4(a) shows a LEED pattern measured at 137 V on $Al_2O_3(0001)$ which has been hydrogen etched at 1180 °C for 5 minutes at 750 mbar and subsequently annealed in ultra-high vacuum (UHV) at 1200 °C for 15 minutes. To heat the sapphire, we placed the sample onto a piece of semiconducting 6H-SiC which is resistively heated. We point out that the sapphire was etched and that we could get a good LEED signal down to 60 V. Below that voltage, the pattern was strongly deformed by surface charging. Moreover, we superimpose the theoretical pattern to the data and the matching is essentially perfect. This means that, at this energy, the path of the diffracted electrons is not affected by surface charging. The LEED analysis performed on graphene grown on pristine sapphire is reported in Fig. S4(b). In this case the $(\sqrt{31} \times \sqrt{31})R \pm 9$ is faintly visible. A much higher background signal as well as a blurring of the diffraction spots originating from the reconstruction are present. This diffuse signal is due to a larger disorder present in this sort of samples. It is not surprising that the reconstruction develops even though, the absence of a well-defined step-terrace structure, limits the extent of the reconstruction domain and therefore the domain-size of graphene. This contributes to a generally more disordered interface in this kind of samples.

In Fig. S4(c) we display the scanning tunnelling spectroscopy acquired on the sample at 78 K. The Dirac point is found 85 meV away from the Fermi level. If we assume a Fermi velocity of $1.1 \times 10^6$ m/s, we retrieve a carrier concentration of about $5 \times 10^{11}$ cm$^{-2}$. In panel (d) we show a portion of the LEED pattern as shown in Fig. 2 (a) of the main text, zoomed-in in the vicinity of the graphene (01) and sapphire (11) reflections. In addition to the $(\sqrt{31} \times \sqrt{31})R \pm 9$ spots (light grey and green, respectively), sapphire (black) and graphene R30 (red), we added the graphene R28 (green) and R32 (blue) to it, to show the coincidence between those orientation and the sapphire reconstructions. The graphene lattice parameter in this example is 1% strained with respect to the 2.461 Å relaxed lattice value.

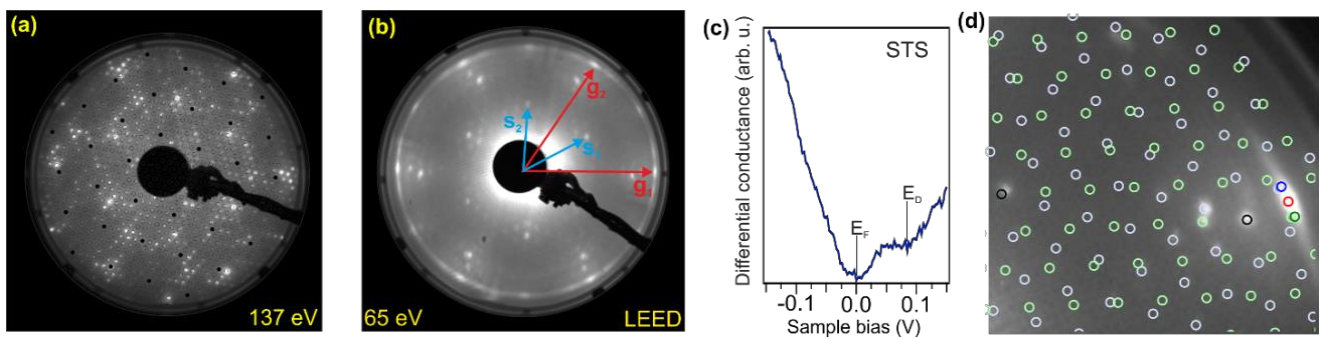

Figure S4. (a) LEED pattern on H$_2$-etched surface of sapphire substrate (before graphene growth). (b) LEED pattern on graphene grown on pristine (non H$_2$-etched) sapphire substrate. (c) Scanning tunnelling spectroscopy data acquired on Gr/Al/Al$_2$O$_3$ at 78 K. (d) zoom-in of the top-right region of Fig. 2 (a) of the main text with superimposed the kinematic

LEED pattern. $(\sqrt{31} \times \sqrt{31})R \pm 9$ spots are light grey and green, respectively. Sapphire (black), graphene R30 (red), we added the graphene R28 (green) and R32 (blue).

## LEEM analysis

Figure S5 shows LEEM micrographs in bright (a) and dark (b) field. The dark field LEEM yields contrast by imaging the electrons coming from a specific diffraction spot, in this case the R30 graphene. What appears in lighter grey in the bright field image corresponds to the R30-graphene, as indicated by the dark field image. From these images, we can also give an estimation about the single-crystal grain size of graphene, i.e. about 1 µm. Indeed, performing µLEED on a single bright-region domain with reference to the inset of panel (a), a (√31×√31) R9 pattern is recognizable, together with a single R30 graphene pattern. This is a clear indication of the fact that, on one hand graphene grows with a high degree of crystallinity on the fully reconstructed sapphire surface and, on the other hand, the graphene single-crystal domain is essentially limited by the grain size of the reconstructed domains on the $Al_2O_3$(0001) surface.

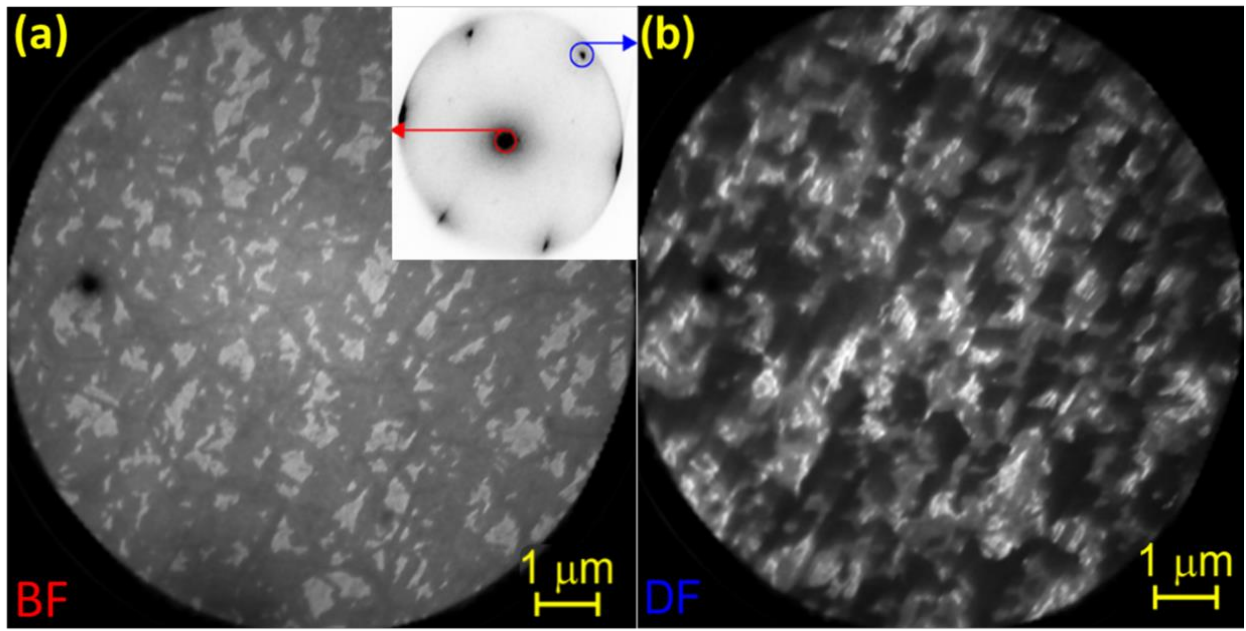

Figure S5: LEEM micrographs recorded on graphene on sapphire in bright field (a) and dark field (b). The diffraction spot selected for the dark field is that of graphene R30, as indicated in the µLEED pattern shown in panel (a).

# Characterization of Graphene Growth over 4-inch and 6-inch wafers

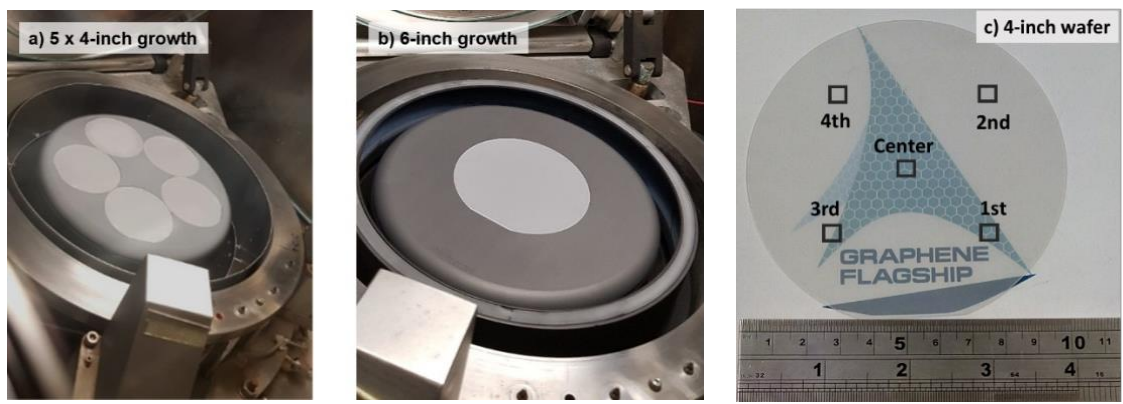

d)

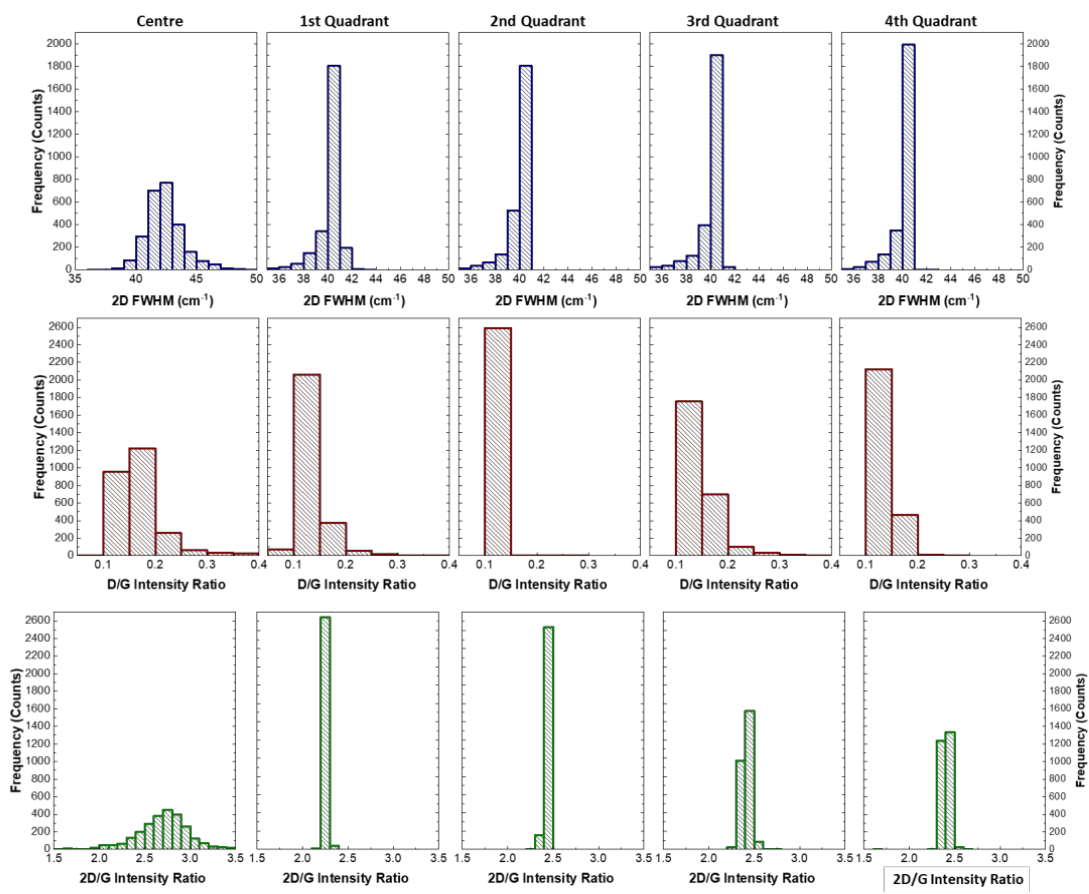

Figure S6: Growth reactor (AIXTRON CCS 2D) used for production of graphene onto (a) 5 x 4-inch and (b) 1 x 6-inch sapphire wafers. (c) Graphene on 4-inch sapphire wafer. (d) Detailed Raman maps of the 2D FWHM, D/G intensity ratio and 2D/G intensity ratio for the different areas in analysis of the 4-inch wafer.

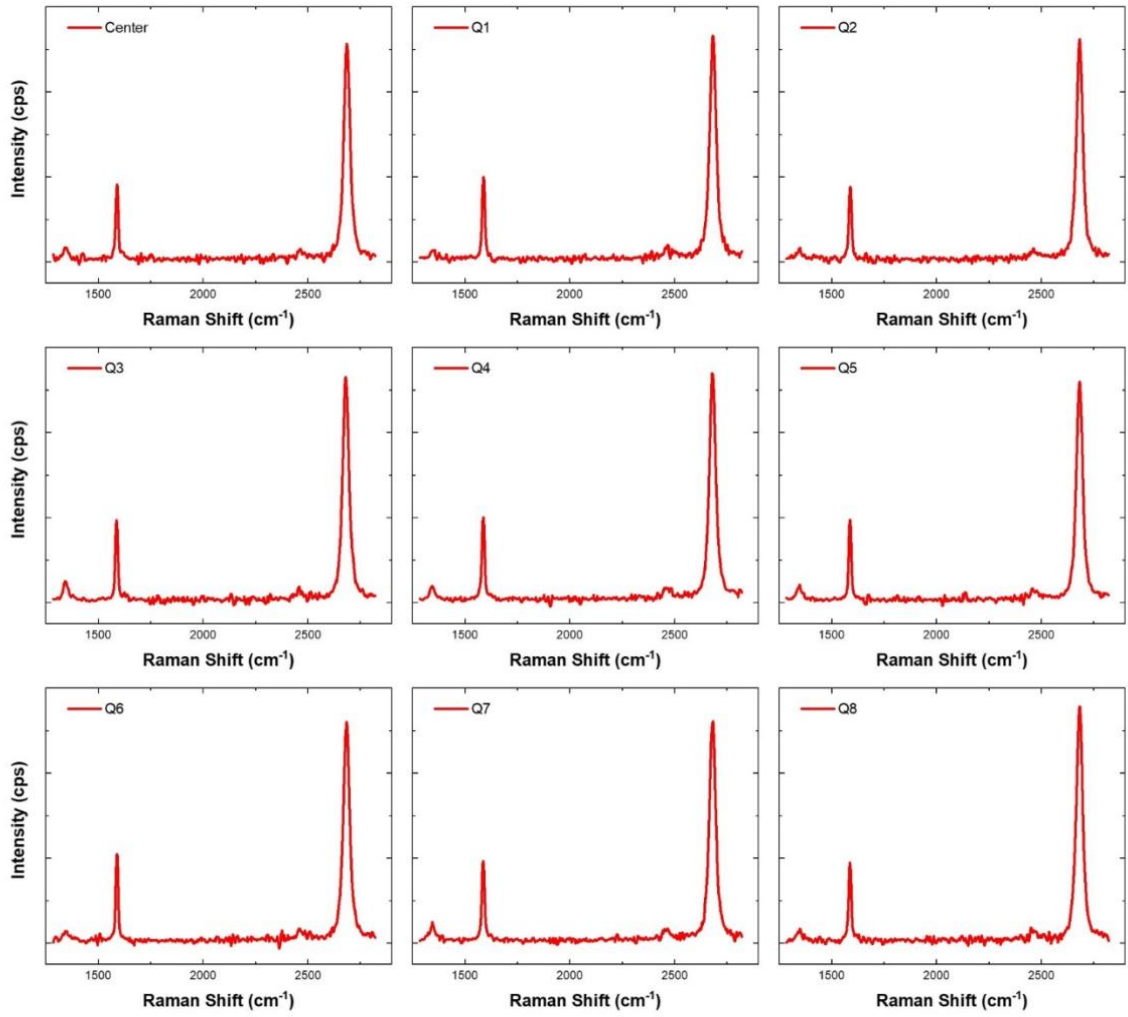

Figure S7. Representative Raman spectra taken in 9 quadrants of the 6 inch wafer (Fig. 3 in the main text).

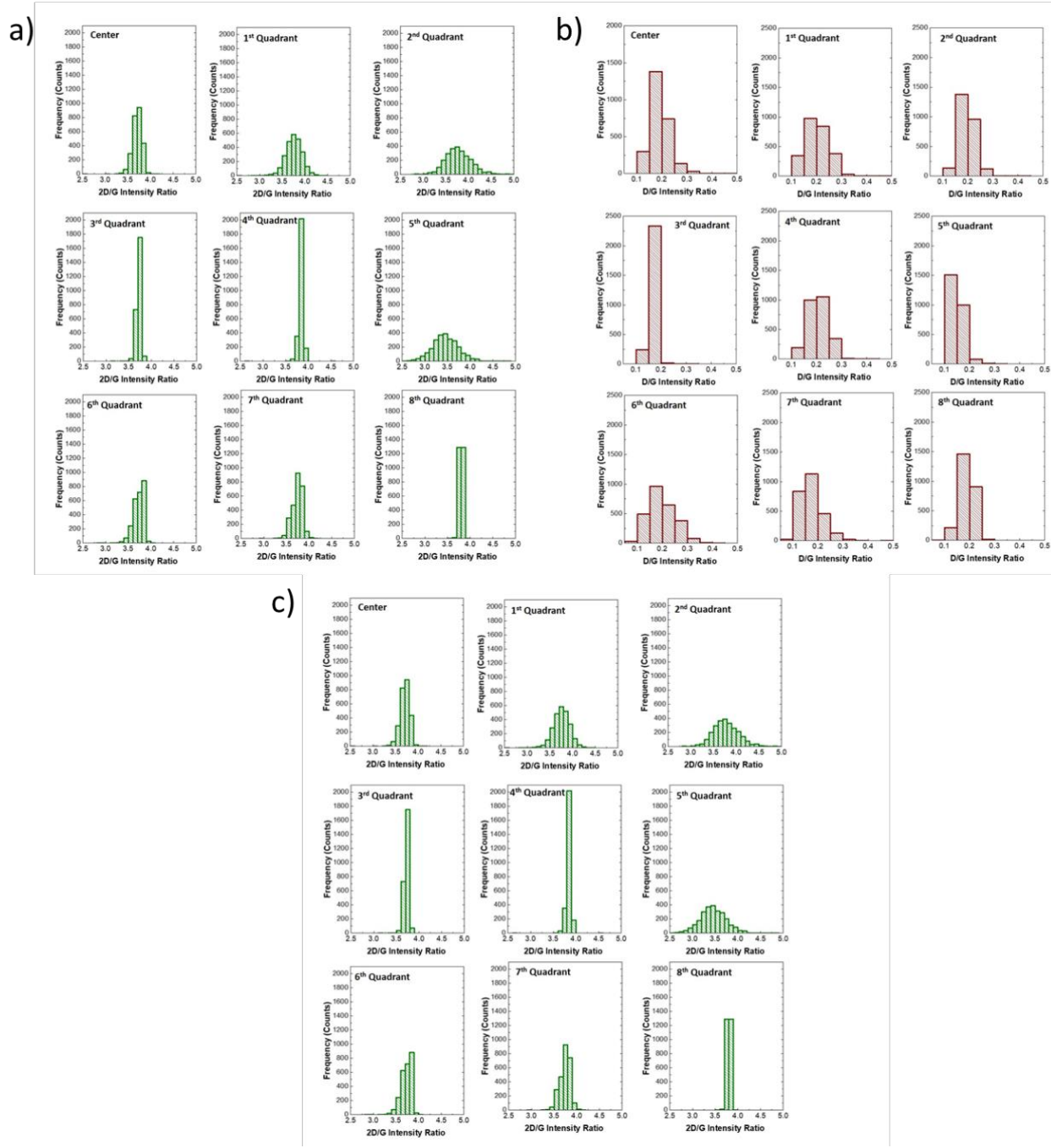

Figure S8: Raman measurements at 9 points across the center and radius of a 6-inch wafer (see spots indicated in Fig. 3(a) of the main text). Detailed Raman maps of the a) 2D FWHM, b) D/G intensity ratio and c) 2D/G intensity ratio for the different areas in analysis.

| Quadrant | 2D/G | 2D FHWM | D/G |
|---|---|---|---|
| Centro | 3.70±0.25 | 36.5±2.1 | 0.19±0.12 |
| Q1 | 3.75±0.42 | 36.0±1.0 | 0.20±0.10 |
| Q2 | 3.78±0.49 | 35.5±2.3 | 0.20±0.08 |
| Q3 | 3.78±0.30 | 36.5±3.0 | 0.17±0.10 |
| Q4 | 3.83±0.35 | 35.8±2.5 | 0.20±0.11 |
| Q5 | 3.45±0.30 | 36.5±4.2 | 0.15±0.06 |
| Q6 | 3.73±0.30 | 36.3±2.5 | 0.2±0.12 |
| Q7 | 3.74±0.26 | 36.0±4.0 | 0.17±0.11 |
| Q8 | 3.80±0.13 | 35.5±1.0 | 0.18±0.07 |

Table S2. Average value and half width at half maximum of the Raman histograms of 2D/G, 2D FWHM, D/G obtained in the 9 different quadrants indicated in Figure 3 in the main text.

**Terahertz time-domain spectroscopy (THz-TDS):**

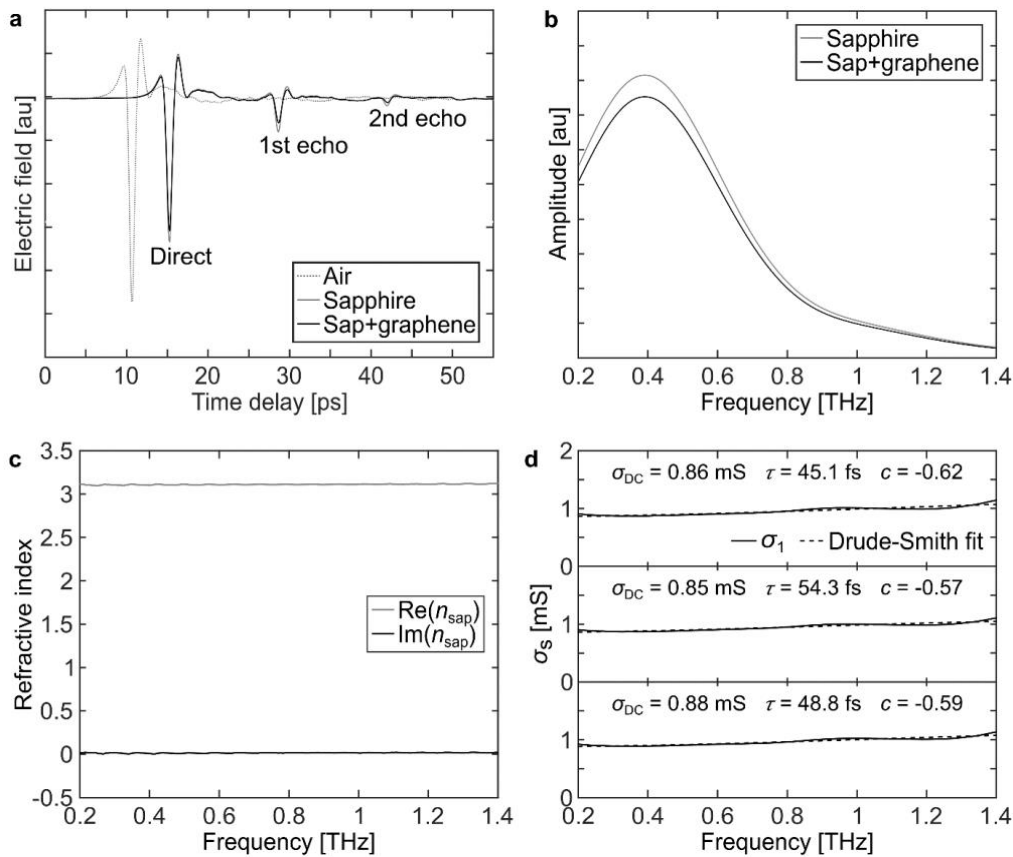

Figure S9. (a) Examples of THz-TDS time-domain waveforms when transmitting through air, a ~640 μm thick sapphire wafer, and a graphene-covered sapphire wafer. The waveforms from sapphire and graphene-covered sapphire shows multiple echoes from partial internal reflections in the sapphire wafer. (b) FFT amplitude from the time-windowed signal of the directly transmitted transients of the waveforms shown in (a). (c) Refractive index of the sapphire wafer calculated from the waveforms from the bare sapphire wafer relative to air. The values of Im($n_{sap}$) are in the 0.005-0.01 range. (d) Examples of representative sheet conductivity spectra for graphene on sapphire together with fits to the Drude-Smith model.

**Wafer-scale polymer-assisted transfer**

Graphene was transferred from sapphire to a target substrate (in this case 90 nm $SiO_2$ on Si) by a peeling approach[2,3]. Samples of graphene on sapphire were submerged in DI water at room temperature for at least 3 hours. A commercial polyvinyl alcohol (PVA) polymer foil was subsequently laminated onto the graphene/sapphire sample and the PVA foil was used to gently peel the graphene sheet from the sapphire substrate[4]. The produced graphene/PVA foil was laminated onto the target substrate and the PVA was subsequently dissolved in DI water. Optical maps were created by acquiring images while raster scanning samples. The optical images were acquired using a Nikon Eclipse L200N microscope equipped with a programmable Prior Scientific XYZ stage. The images were stitched together to form an optical map after background subtraction[5].

**FET measurements:**

The linear fitting of the conductance versus gate voltage, used in the main text to obtain the mobility, ignores the presence of any carrier-independent contribution to the mobility, such as contact resistance or short-range scattering. To address the presence of those effects, we also attempt the method of fitting the transfer curve[1], as shown in Fig.S10. From the fitting we obtain values of the mobility of $\mu_e$ = 1900 cm$^2$/Vs $\mu_h$ = 2400 cm$^2$/Vs for electron and holes respectively, in agreement with what we report in the text. To evaluate the quality of graphene we also evaluate n*, as the potential fluctuation of electrons in graphene[5]. This parameter can be determined by fitting log(σ) versus log(n) in the constant (low n) and linear (high n) regime[6]

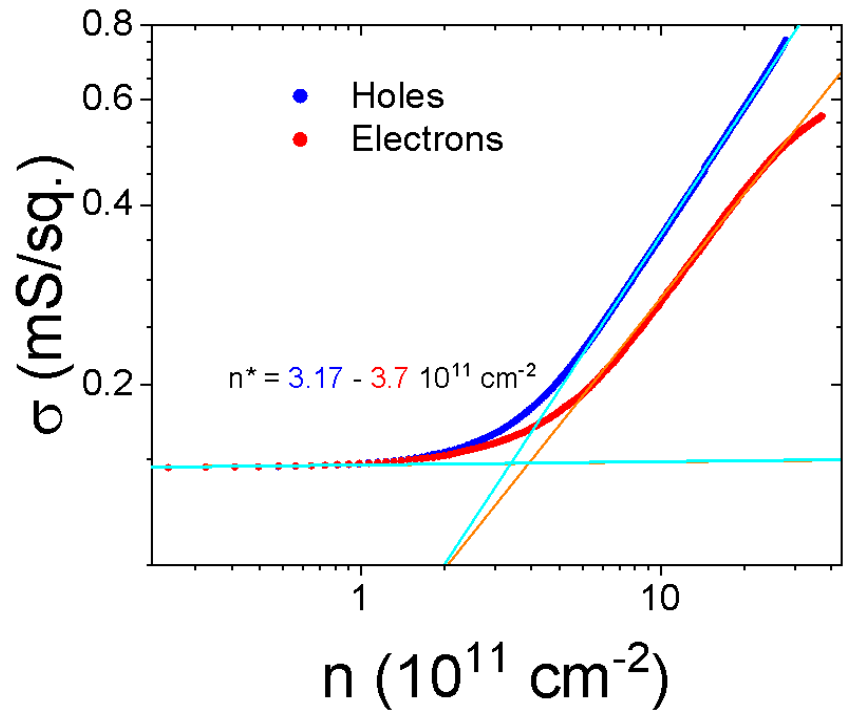

Figure S10. Procedure to extract n*: two lines are used to fit the double logarithmic plot of σ as function of n. We obtain n* around $3\times10^{11}$cm$^{-2}$, with a slight difference in the conductance with a small difference between electrons (black) and holes (red).

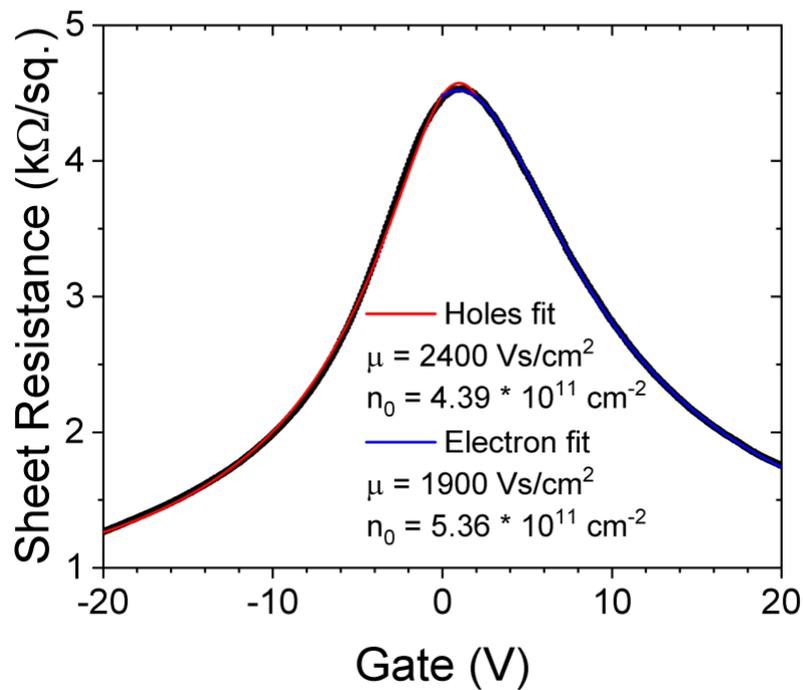

Figure S11. Transfer curve of a graphene device on SiO$_2$/Si substrate: we used two different set of fitting parameters for the electrons and holes majority state, to address the asymmetry of the curve. The values obtained are in good agreement with the main text.

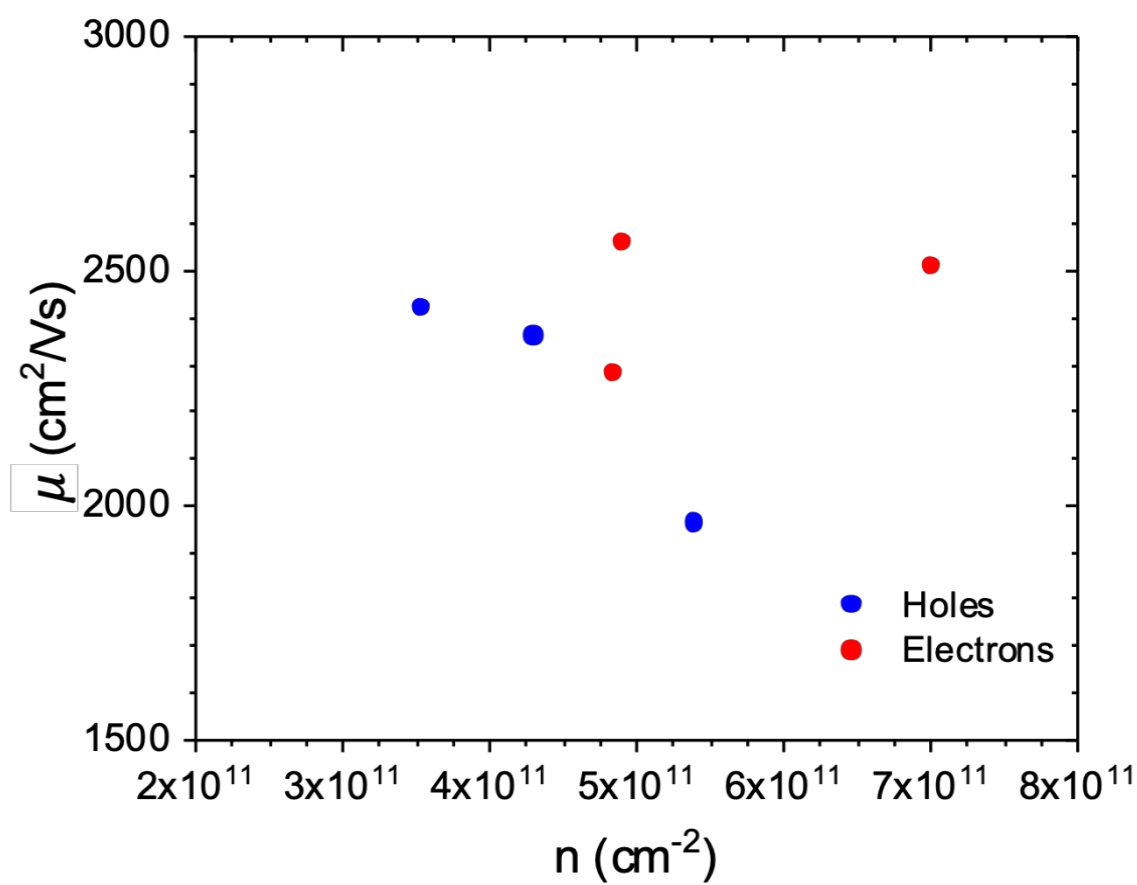

Figure S12. Mobility-vs-carrier density statistic on different graphene devices on $SiO_2$/Si substrate.

**TXRF measurements**

TXRF measurements were carried out (by EAG Laboratories, Sunnyvale, USA) on 4-inch sapphire wafers before and after growth of graphene. A tungsten (W) rotating anode was used as the source, with measurements taken at 49 points on each 4-inch wafer. Table S2 shows the average detected concentration of elements on each wafer. We observe that after graphene growth, metal contamination levels still satisfy back-end-of-line (BEOL) integration requirements (metal below $10^{12}$ at/$cm^2$). Indeed, for most analyzed elements, there is a general improvement (i.e. lower values) upon graphene growth. This is most likely to due to the fact that starting sapphire wafers, being manually handled and not within dedicated fab premises, already presented contaminants which diffused into the sapphire during the high-temperature growth process. We expect that the starting contaminants would be reduced by automated loading of wafers in a clean, dedicated environment, and are confident that with such procedures, front-end-of-line (FEOL) requirements, $10^{10}$ at/$cm^2$, would be met.

| | Si | P | S | Cl | K | Ca | Ti | Cr | Mn | Fe | Ni | Cu | Zn |
|---|---|---|---|---|---|---|---|---|---|---|---|---|---|
| Sapphire Reference | 2.86 | 24.91 | 98.53 | 102.51 | 9.53 | 78.52 | 0.63 | 4.4 | 1.31 | 44.61 | 1.71 | 5.75 | 3.21 |
| Graphene on sapphire | 6.42 | 3.3 | 60.58 | 60.69 | 19.14 | 36.71 | 5.24 | 0.83 | 0.35 | 17.05 | 0.05 | 0.29 | 6.28 |

Table S3: TXRF results for 4" sapphire wafers with and without graphene growth. The average concentration over 49 measurement points are shown (units of $10^{10}$ atoms/cm$^2$).

# References

1. Yang, S. Y. *et al.* Metal-Etching-Free Direct Delamination and Transfer of Single-Layer Graphene with a High Degree of Freedom. *Small* **11**, 175–181 (2015).

2. Whelan, P. R. *et al.* Raman spectral indicators of catalyst decoupling for transfer of CVD grown 2D materials. *Carbon N. Y.* **117**, 75–81 (2017).

3. Shivayogimath, A. *et al.* Do-It-Yourself Transfer of Large-Area Graphene Using an Office Laminator and Water. *Chem. Mater.* **31**, 2328–2336 (2019).

4. Jessen, B. S. *et al.* Quantitative optical mapping of two-dimensional materials. *Sci. Rep.* **8**, 6381 (2018).

5. Couto, N. J. G. *et al.* Random Strain Fluctuations as Dominant Disorder Source for High-Quality On-Substrate Graphene Devices. *Phys. Rev. X* **4**, 041019 (2014).

6. Banszerus, L. *et al.* Ultrahigh-mobility graphene devices from chemical vapor deposition on reusable copper. *Sci. Adv.* **1**, e1500222 (2015).